\documentclass{amsart}
\usepackage{amssymb}
\usepackage{xcolor,graphicx}
\usepackage{bbm}

\theoremstyle{plain}
\newtheorem{theorem}{Theorem}
\newtheorem{lemma}[theorem]{Lemma}
\newtheorem{corollary}[theorem]{Corollary}
\theoremstyle{definition}

\theoremstyle{remark}

\providecommand{\abs}[1]{\lvert#1\rvert}
\providecommand{\Abs}[1]{\Bigl\lvert#1\Bigr\rvert}
\providecommand{\norm}[1]{\lVert#1\rVert}

\begin{document}

\title[Bayesian nonparametrics on a Fr\'echet class]{Bayesian nonparametric inference\\on a Fr\'echet class}

\author{Emanuela Dreassi}
\address{Emanuela Dreassi, Dipartimento di Statistica, Informatica, Applicazioni ``G. Parenti'', Universit\`a di Firenze, viale Morgagni 59, 50134 Firenze, Italy}
\email{emanuela.dreassi@unifi.it}
\author{Luca Pratelli}
\address{Luca Pratelli, Accademia Navale, viale Italia 72, 57100 Livorno,
Italy} \email{luca{\_}pratelli@marina.difesa.it}
\author{Pietro Rigo}
\address{Pietro Rigo (corresponding author), Dipartimento di Scienze Statistiche ``P. Fortunati'', Universit\`a di Bologna, via delle Belle Arti 41, 40126 Bologna, Italy}
\email{pietro.rigo@unibo.it}

\begin{abstract}
Let $(\mathcal{X},\mathcal{F},\mu)$ and $(\mathcal{Y},\mathcal{G},\nu)$ be probability spaces and $(Z_n)$ a sequence of random variables with values in $(\mathcal{X}\times\mathcal{Y},\,\mathcal{F}\otimes\mathcal{G})$. Let $\Gamma(\mu,\nu)$ be the collection of all probability measures $p$ on $\mathcal{F}\otimes\mathcal{G}$ such that
$$p\bigl(A\times\mathcal{Y}\bigr)=\mu(A)\quad\text{and}\quad p\bigl(\mathcal{X}\times B\bigr)=\nu(B)\quad\text{for all }A\in\mathcal{F}\text{ and }B\in\mathcal{G}.$$
In this paper, we build some probability measures $\Pi$ on $\Gamma(\mu,\nu)$. In addition, for each such $\Pi$, we assume that $(Z_n)$ is exchangeable with de Finetti's measure $\Pi$ and we evaluate the conditional distribution $\Pi(\cdot\mid Z_1,\ldots,Z_n)$. In Bayesian nonparametrics, if $(Z_1,\ldots,Z_n)$ are the available data, $\Pi$ and $\Pi(\cdot\mid Z_1,\ldots,Z_n)$ can be regarded as the prior and the posterior, respectively. To support this interpretation, it suffices to think of a problem where the unknown probability distribution of some bivariate phenomenon is constrained to have marginals $\mu$ and $\nu$. Finally, analogous results are obtained for the set $\Gamma(\mu)$ of those probability measures on $\mathcal{F}\otimes\mathcal{G}$ with marginal $\mu$ on $\mathcal{F}$ (but arbitrary marginal on $\mathcal{G}$). That is, we introduce some priors on $\Gamma(\mu)$ and we evaluate the corresponding posteriors.
\end{abstract}

\maketitle

\section{Introduction}\label{intro}

\noindent An intriguing problem in Bayesian nonparametrics is to build a (suitable) prior distribution supported by a given subset $\mathcal{P}_0$ of the set of all probability measures on the sample space. Depending on $\mathcal{P}_0$, such a problem may be quite hard. This paper focus on this problem when $\mathcal{P}_0$ is a Fr\'echet class (to be meant as in Section \ref{v5th7u8}).

\medskip

\noindent Throughout, $(\mathcal{Z},\mathcal{H})$ is a measurable space and $(Z_n:n\ge 1)$ a sequence of random variables with values in $(\mathcal{Z},\mathcal{H})$. All random elements appearing in this paper are defined on the same probability space, say $(\Omega,\mathcal{A},\mathbb{P})$. Moreover, $\mathcal{P}$ denotes the collection of all probability measures on $\mathcal{H}$ and $\Psi$ the $\sigma$-field over $\mathcal{P}$ generated by the evaluation maps $p\mapsto p(H)$ for all $H\in\mathcal{H}$.

\medskip

\subsection{Bayesian nonparametric inference with exchangeable data} In a nutshell, the standard framework of Bayesian nonparametrics can be summarized as follows.

\medskip

\noindent Let $Z$ be a random element, taking values in $(\mathcal{Z},\mathcal{H})$, and let $\mathcal{L}(Z)$ denote its probability distribution. The goal is to make inference on $\mathcal{L}(Z)$, or on some functional of $\mathcal{L}(Z)$, based on the available observations $Z_1,Z_2,\ldots$ on $Z$. To this end, some assumptions on the data sequence $(Z_n)$ are to be made. It is quite usual, even if not mandatory, to assume $(Z_n)$ exchangeable or partially exchangeable; see e.g. \cite{ANTONIAK}, \cite{BPR13}, \cite{BDPR21}, \cite{STS25}, \cite{SONIA}, \cite{GHOSALVDV}. In the sequel, $(Z_n)$ is supposed to be exchangeable. We refer to \cite{ALDOUS} for an exhaustive treatment of exchangeability.

\medskip

\noindent Since $(Z_n)$ is exchangeable, under mild conditions on $(\mathcal{Z},\mathcal{H})$, there is a unique probability measure $\Pi$ on $\Psi$ such that
$$\mathbb{P}\bigl(Z_1\in H_1,\ldots,Z_k\in H_k\bigr)=\int p(H_1)\ldots p(H_k)\,\Pi(dp)$$
for all $k\ge 1$ and $H_1,\ldots,H_k\in\mathcal{H}$. Such a $\Pi$ is called the {\em de Finetti's measure} or the {\em prior distribution}. Moreover, for each $n\ge 1$, the sequence $(Z_{n+k}:k\ge 1)$ is still exchangeable conditionally on $(Z_1,\ldots,Z_n)$. Hence, there is an a.s. unique probability measure $\Pi(\cdot\mid Z_1,\ldots,Z_n)$ on $\Psi$ such that
$$\mathbb{P}\bigl(Z_{n+1}\in H_1,\ldots,Z_{n+k}\in H_k\mid Z_1,\ldots,Z_n\bigr)=\int p(H_1)\ldots p(H_k)\,\Pi(dp\mid Z_1,\ldots,Z_n)$$
a.s. for all $k\ge 1$ and $H_1,\ldots,H_k\in\mathcal{H}$. If $(Z_1,\ldots,Z_n)$ are the available observations, $\Pi(\cdot\mid Z_1,\ldots,Z_n)$ is called the {\em posterior distribution}.

\medskip

\noindent In real problems, to make inference on $\mathcal{L}(Z)$, one first selects a prior $\Pi$ and then evaluates the posterior $\Pi(\cdot\mid Z_1,\ldots,Z_n)$. Clearly, evaluating the posterior may be very difficult, and one is often forced to use numerical approximations. Anyway, if and when $\Pi(\cdot\mid Z_1,\ldots,Z_n)$ is available, any inference regarding $\mathcal{L}(Z)$ should be based on it. Among other things, given $\Pi(\cdot\mid Z_1,\ldots,Z_n)$, one can evaluate the {\em predictive distribution}
$$\mathbb{P}(Z_{n+1}\in H\mid Z_1,\ldots,Z_n)=\int p(H)\,\Pi(dp\mid Z_1,\ldots,Z_n)\quad\quad\text{for all }H\in\mathcal{H}.$$

\medskip

\noindent Suppose now that $\mathcal{L}(Z)$ is known to belong to some class $\mathcal{P}_0\in\Psi$. This actually happens in a number of situations. Some obvious examples are:

\medskip

\begin{itemize}

\item $\mathcal{P}_0=\bigl\{p\in\mathcal{P}:p\text{ is absolutely continuous with respect to }\lambda\bigr\}$

\noindent where $\lambda$ is a reference measure on $\mathcal{H}$; see \cite{BPR13};

\medskip

\item $\mathcal{P}_0=\bigl\{p\in\mathcal{P}:p\text{ is invariant under }G\bigr\}$

\noindent where $G$ is a group of transformations on $\mathcal{Z}$; see \cite[Example 17]{BDLPR23}, \cite{DALAL}, \cite{HZ};

\medskip

\item $\mathcal{P}_0=\bigl\{p\in\mathcal{P}:p=p^*\text{ on }\mathcal{H}^*\bigr\}$

\noindent where $\mathcal{H}^*\subset\mathcal{H}$ and $p^*$ is a given element of $\mathcal{P}$;

\medskip

\item $\mathcal{Z}=\mathbb{R}$ and $\mathcal{P}_0=\bigl\{p\in\mathcal{P}:p\text{ has given moments up to a certain order}\bigr\}$.

\end{itemize}

\medskip

\noindent Obviously, if it is known that $\mathcal{L}(Z)\in\mathcal{P}_0$, the prior $\Pi$ should satisfy
$$\Pi(\mathcal{P}_0)=1.$$
However, to build a reasonable prior, with large support and satisfying $\Pi(\mathcal{P}_0)=1$, is usually hard; see e.g. \cite{DALAL}. Similarly, once a prior is selected, to obtain the posterior may be difficult as well.

\medskip

\subsection{Fr\'echet classes}\label{v5th7u8} In the sequel,
$$(\mathcal{X},\mathcal{F},\mu)\quad\text{and}\quad (\mathcal{Y},\mathcal{G},\nu)$$
are probability spaces. To avoid annoying complications, $(\mathcal{X},\mathcal{F})$ and $(\mathcal{Y},\mathcal{G})$ are assumed to be {\em standard Borel spaces}.

\medskip

\noindent Let
$$\mathcal{Z}=\mathcal{X}\times\mathcal{Y}\quad\text{and}\quad\mathcal{H}=\mathcal{F}\otimes\mathcal{G}.$$
For each $p\in\mathcal{P}$, the marginals of $p$ are
$$p_1(A)=p\bigl(A\times\mathcal{Y}\bigr)\quad\text{and}\quad p_2(B)=p\bigl(\mathcal{X}\times B\bigr)\quad\text{for all }A\in\mathcal{F}\text{ and }B\in\mathcal{G}.$$
Obviously, $p_1$ and $p_2$ are probability measures on $\mathcal{F}$ and $\mathcal{G}$, respectively. The {\em Fr\'echet class} of the pair $(\mu,\nu)$ is
$$\Gamma(\mu,\nu)=\bigl\{p\in\mathcal{P}:p\text{ has marginal }\mu\text{ on }\mathcal{F}\text{ and }\nu\text{ on }\mathcal{G}\bigr\}.$$
Similarly, in this paper,
$$\Gamma(\mu)=\bigl\{p\in\mathcal{P}:p\text{ has marginal }\mu\text{ on }\mathcal{F}\bigr\}$$
is said to be the Fr\'echet class of $\mu$.

\medskip

\noindent Our goal is to build some priors $\Pi$ such that $\Pi(\mathcal{P}_0)=1$ where
$$\mathcal{P}_0=\Gamma(\mu,\nu)\quad\text{or}\quad\mathcal{P}_0=\Gamma(\mu).$$
In addition, assuming to have an exchangeable sequence $(Z_n)$ with one of such priors, we evaluate the corresponding posterior.

\medskip

\noindent The first choice of $\mathcal{P}_0$ does not require a lot of comments. Indeed, $\Gamma(\mu,\nu)$ plays a role in a number of settings, including contingency tables, mass transportation, Wasserstein distances, statistical physics, gradient flows and their many ramifications; see e.g. \cite{AGS}, \cite{KDO} and references therein. We just note that $\Gamma(\mu,\nu)$ reduces to the collection of bivariate copulas if $\mathcal{X}=\mathcal{Y}=[0,1]$ and $\mu$ and $\nu$ are uniform. Hence, in this case, our problem is to build a (reasonable) prior on the set of bivariate copulas. The second choice of $\mathcal{P}_0$, while less intriguing than the first, makes some interest as well. In general, $\Gamma(\mu)$ comes into play if one marginal of $\mathcal{L}(Z)$ is known but the other is not. As an example, suppose $Z=(X,Y)$ where $Y$ is a response variable and $X$ a vector of covariates. If $\mathcal{L}(X)$ is known, say $\mathcal{L}(X)=\mu$, making inference on $\mathcal{L}(Z)$ amounts to making inference on the conditional distribution $\mathcal{L}(Y\mid X)$. Hence, for instance, $\Gamma(\mu)$ is meaningful when $\mathcal{L}(X)$ is known and the object of inference is some functional of $\mathcal{L}(Y\mid X)$, such as the regression function $g(x)=E(Y\mid X=x)$. Note also that there are situations where $\mathcal{L}(X)$ is given while $\mathcal{L}(Y\mid X)$ is arbitrary. A well known example, concerning variable selection, is the knockoff procedure by Barber and Candes; see e.g. \cite{BC15} and \cite{CFJL18}.

\medskip

\subsection{Content of this paper} As already noted, we aim to introduce some  priors, supported by $\Gamma(\mu,\nu)$ or $\Gamma(\mu)$, and to evaluate the corresponding posteriors.

\medskip

\noindent To fix ideas, let us consider $\Gamma(\mu,\nu)$ (the case $\Gamma(\mu)$ is quite analogous). A prior on $\Gamma(\mu,\nu)$ can be regarded as the probability distribution of a random probability measure (see Section \ref{b78m2}) taking values in $\Gamma(\mu,\nu)$. Accordingly, each of our priors is the probability distribution of a measurable map $P:\Omega\rightarrow\Gamma(\mu,\nu)$, and our main task is to build such a $P$. In Section \ref{ga332c}, in addition to $P\in\Gamma(\mu,\nu)$, we require $P$ to admit a density with respect to the product measure $\mu\times\nu$. This has two advantages. First, at least for us, it is easier to argue in terms of densities than in terms of probability measures. Second, if $P$ is absolutely continuous with respect $\mu\times\nu$, we are actually dealing with a dominated statistical model. Hence, the posterior can be evaluated via Bayes theorem.

\medskip

\noindent Once a random probability measure $P$ has been built, we look for explicit formulas for its probability distribution. Our results are more or less informative depending on the different situations. In some cases, we completely obtain the finite dimensional distributions of $P$, that is, we evaluate the distribution of the random vector $\bigl(P(H_1),\ldots,P(H_m)\bigr)$ where $H_1,\ldots,H_m$ is a finite measurable partition of $\mathcal{X}\times\mathcal{Y}$. In other cases, we only find the distribution of $P(H)$ for fixed $H\in\mathcal{H}$. Finally, as regards the posterior, we obtain the conditional distribution of $\bigl(P(H_1),\ldots,P(H_m)\bigr)$ or $P(H)$ given $(Z_1,\ldots,Z_n)$.

\medskip

\section{Random probability measures}\label{b78m2}

\noindent In this section, we just report some well known definitions. Recall that $(\Omega,\mathcal{A},\mathbb{P})$ is a probability space and $(\mathcal{Z},\mathcal{H})$ a measurable space. Moreover, $\mathcal{P}$ is the set of all probability measures on $\mathcal{H}$ and $\Psi$ the $\sigma$-field over $\mathcal{P}$ generated by the evaluation maps $p\mapsto p(H)$ for all $H\in\mathcal{H}$.

\medskip

\noindent A {\em random probability measure} (r.p.m.) on $(\mathcal{Z},\mathcal{H})$ is a measurable map
$$P:(\Omega,\mathcal{A})\rightarrow (\mathcal{P},\Psi).$$
Equivalently, a r.p.m. is a map $P:\Omega\rightarrow\mathcal{P}$ such that $\omega\mapsto P(\omega,H)$ is a real random variable for fixed $H\in\mathcal{H}$. Obviously, the distribution of a r.p.m. $P$ is the probability measure $\Pi$ on $\Psi$ defined as
$$\Pi(L)=\mathbb{P}(P\in L)\quad\quad\text{for }L\in\Psi.$$
In the sequel, $P(H)$ denotes the real random variable $\omega\mapsto P(\omega,H)$.

\medskip

\noindent Let $\alpha_1,\ldots,\alpha_m$ be non-negative constants such that $\sum_{i=1}^m\alpha_i>0$ and let $U_1,\ldots,U_m$ be independent Gamma random variables with scale parameter 1 and shape parameter $\alpha_i$ (with $U_i=0$ a.s. if $\alpha_i=0$). The Dirichlet law of parameters $\alpha_1,\ldots,\alpha_m$, denoted
$$\text{Dir}(\alpha_1,\ldots,\alpha_m),$$
is the probability distribution of the vector
$$\left(\frac{U_1}{\sum_{i=1}^mU_i},\ldots,\frac{U_m}{\sum_{i=1}^mU_i}\right).$$

\medskip

\noindent Let $c>0$ be a constant and $\lambda\in\mathcal{P}$. A  r.p.m. $P$ is said to be Dirichlet with parameters $c$ and $\lambda$ if
$$\Bigl(P(H_1),\ldots,P(H_m)\Bigr)\,\sim\,\text{Dir}\bigl[c\,\lambda(H_1),\ldots,c\,\lambda(H_m)\bigr]$$
whenever $H_1,\ldots,H_m$ is a finite measurable partition of $\mathcal{Z}$. A Dirichlet prior is the probability distribution of a Dirichlet r.p.m. Dirichlet priors are a basic ingredient of Bayesian nonparametrics. They also play a role in many other frameworks, including population genetics and species sampling. Here, we just recall their coniugacy property. Suppose $(Z_n)$ is exchangeable and the prior $\Pi$ is Dirichlet with parameters $c$ and $\lambda$. Then, the posterior $\Pi(\cdot\mid Z_1,\ldots,Z_n)$ is still Dirichlet with parameters $c+n$ and $\frac{c\lambda+\sum_{i=1}^n\delta_{Z_i}}{c+n}$, where $\delta_z$ denotes the unit mass at $z$ for every $z\in\mathcal{Z}$. See e.g. \cite{BLAKMQ}, \cite{FERG}, \cite{SET}.

\medskip

\noindent Despite their various merits, however, Dirichlet priors do not work for our purposes. Suppose in fact $\mathcal{Z}=\mathcal{X}\times\mathcal{Y}$, $\mathcal{H}=\mathcal{F}\otimes\mathcal{G}$, and $\Pi$ is a Dirichlet prior. The parameter $\lambda$ of $\Pi$ is a probability measure on $\mathcal{F}\otimes\mathcal{G}$. If $\lambda$ is not trivial on $\mathcal{F}$, then
$$\Pi\bigl(\Gamma(\mu,\nu)\bigr)=\Pi\bigl(\Gamma(\mu)\bigr)=0.$$
To see this, take a Dirichlet r.p.m. $P$ with distribution $\Pi$. Since $\lambda$ is not trivial on $\mathcal{F}$, there is $A\in\mathcal{F}$ such that $\lambda(A\times\mathcal{Y})\in (0,1)$. Hence, $P(A\times\mathcal{Y})$ has a beta distribution, which in turn implies
$$\Pi\bigl(\Gamma(\mu,\nu)\bigr)\le\Pi\bigl(\Gamma(\mu)\bigr)=\mathbb{P}\bigl(P\in\Gamma(\mu)\bigr)\le\mathbb{P}\bigl[P(A\times\mathcal{Y})=\mu(A)\bigr]=0.$$

\medskip

\noindent Finally, a {\em copula} is a distribution function on $\mathbb{R}^n$ whose marginals are uniform on $[0,1]$. If $C$ is a copula and $F_1,\ldots,F_n$ are distribution functions on $\mathbb{R}$, then
\begin{gather}\label{n8u}
F(x_1,\ldots,x_n)=C\bigl[F_1(x_1),\ldots,F_n(x_n)\bigr],\quad (x_1,\ldots,x_n)\in\mathbb{R}^n,
\end{gather}
is a distribution function on $\mathbb{R}^n$ with marginals $F_1,\ldots,F_n$. Conversely, for each distribution function $F$ on $\mathbb{R}^n$ (with marginals $F_1,\ldots,F_n$) there is a copula $C$ satisfying equation \eqref{n8u}. Moreover, such a $C$ is unique whenever $F_1,\ldots,F_n$ are continuous. In the sequel, we are concerned with bivariate copulas (i.e., $n=2$).

\medskip

\section{Some priors on $\Gamma(\mu,\nu)$}\label{ga332c}

\noindent Let again $\mathcal{Z}=\mathcal{X}\times\mathcal{Y}$ and $\mathcal{H}=\mathcal{F}\otimes\mathcal{G}$. In this section, we denote by $\lambda$ the product measure
$$\lambda=\mu\times\nu$$
and we focus on
$$\mathcal{P}_0=\bigl\{p\in\Gamma(\mu,\nu):\,p\text{ is absolutely continuous with respect to }\lambda\bigr\}.$$
Our goal is to introduce some priors $\Pi$ such that $\Pi(\mathcal{P}_0)=1$ and to obtain the corresponding posteriors.

\medskip

\noindent Some more notation is needed. Let
\begin{gather*}
\mathcal{D}=\bigl\{f:\,f\text{ is a probability density with respect to }\lambda\text{ and }P_f\in\Gamma(\mu,\nu)\bigr\}
\end{gather*}
where $P_f$ is the probability measure
$$P_f(H)=\int_H f\,d\lambda\quad\quad\text{for all }H\in\mathcal{H}.$$
The set $\mathcal{D}$ is equipped with the $\sigma$-field $\Psi^*$ generated by the map $f\mapsto P_f$. Precisely, $\Psi^*$ is the collection of events of the form $\bigl\{f\in\mathcal{D}:\,P_f\in L\bigr\}$ for some $L\in\Psi$. A {\em random density} is a random element of $(\mathcal{D},\Psi^*)$. Moreover, we denote by
$$Z_n=(X_n,Y_n),\quad\quad n\ge 1,$$
a sequence of random variables taking values in $(\mathcal{X}\times\mathcal{Y},\,\mathcal{H})$.

\medskip

\noindent One advantage of $\mathcal{P}_0$ is that it is a {\em dominated statistical model}, in the sense that each element of $\mathcal{P}_0$ is absolutely continuous with respect to a fixed probability measure on $\mathcal{H}$. Hence, if the prior is concentrated on $\mathcal{P}_0$, the posterior can be obtained via Bayes Therorem. This is formalized in the next result.

\medskip

\begin{lemma}\label{x4r5y}
$(Z_n)$ is exchangeable and the prior is concentrated on $\mathcal{P}_0$ if and only if there is a random density $f$ such that
\begin{gather}\label{nu7d4}
(Z_n)\mid f\,\text{ is i.i.d. with }\,Z_1\sim f.
\end{gather}
Moreover, if $\Pi$ denotes the probability distribution of $f$, one obtains
$$\Pi(df\mid Z_1,\ldots,Z_n)=\frac{\prod_{i=1}^nf(Z_i)}{\int\prod_{i=1}^n\phi(Z_i)\,\Pi(d\phi)}\,\Pi(df)\quad\quad\text{a.s.}$$
whenever $\int\prod_{i=1}^n\phi(Z_i)\,\Pi(d\phi)\in (0,\infty)$. (Here, $\Pi$ and $\Pi(\cdot\mid Z_1,\ldots,Z_n)$ are probability measures on $(\mathcal{D},\Psi^*)$).
\end{lemma}

\begin{proof}
The ``if" part is trivial. Conversely, since $(\mathcal{X},\mathcal{F})$ and $(\mathcal{Y},\mathcal{G})$ are standard Borel spaces, if $(Z_n)$ is exchangeable there is a r.p.m. $P$ on $(\mathcal{X}\times\mathcal{Y},\,\mathcal{H})$ such that
$$(Z_n)\mid P\,\text{ is i.i.d. with }\,Z_1\sim P.$$
Since the prior is concentrated on $\mathcal{P}_0$, one also obtains $P\in\mathcal{P}_0$ a.s. Hence, to get \eqref{nu7d4}, it suffices to take $f$ as the density of $P$ with respect to $\lambda$. Finally, as noted above, since $\mathcal{P}_0$ is a dominated statistical model, the posterior can be evaluated via Bayes theorem. Precisely, assume condition \eqref{nu7d4}, call $\Pi$ the probability distribution of the random density $f$, and define
$$\Pi^*(df)=\frac{\prod_{i=1}^nf(Z_i)}{\int\prod_{i=1}^n\phi(Z_i)\,\Pi(d\phi)}\,\Pi(df)\,\quad\text{whenever }\,\int\prod_{i=1}^n\phi(Z_i)\,\Pi(d\phi)\in (0,\infty).$$
Such a $\Pi^*$ is a r.p.m. on $(\mathcal{D},\Psi^*)$. Moreover, using Fubini's theorem, it is straightforward to show that
$$\int P_f(H_1)\ldots P_f(H_k)\,\Pi^*(df)=\mathbb{P}\bigl(Z_{n+1}\in H_1,\ldots,Z_{n+k}\in H_k\mid Z_1,\ldots,Z_n\bigr)$$
a.s. for all $k\ge 1$ and $H_1,\ldots,H_k\in\mathcal{H}$. Therefore, $\Pi(\cdot\mid Z_1,\ldots,Z_n)=\Pi^*$ a.s.
\end{proof}

\medskip

\noindent A r.p.m. $P$ on $(\mathcal{X}\times\mathcal{Y},\,\mathcal{H})$ such that $P\in\mathcal{P}_0$ a.s. can be identified with a random density. Similarly, a prior on $\mathcal{P}_0$ can be identified with the probability distribution of a random density. Accordingly, in the rest of this section, we will speak of random densities rather than priors on $\mathcal{P}_0$. Importantly, by Lemma \ref{x4r5y}, once a random density has been defined, the posterior can be written in closed form. Hence, from a Bayesian perspective, it remains only to introduce some (suitable) random densities. We will do it following two different approaches.

\medskip

\subsection{First approach}\label{bew5} Let
$$f(x,y)=1+g(x)\,h(y)\quad\quad\text{for all }(x,y)\in\mathcal{X}\times\mathcal{Y},$$
where
\begin{gather}\label{bg67km}
g:\mathcal{X}\rightarrow\mathbb{R}\text{ and }h:\mathcal{Y}\rightarrow\mathbb{R}\text{ are Borel functions such that}
\\\sup\,\abs{g}\le 1,\quad\sup\,\abs{h}\le 1\quad\text{and}\quad E_\mu(g)=E_\nu(h)=0.\notag
\end{gather}
Then, $f$ is a non-negative Borel function on $\mathcal{X}\times\mathcal{Y}$. In addition, for all $A\in\mathcal{F}$ and $B\in\mathcal{G}$, Fubini's theorem yields
\begin{gather*}
\int_{A\times\mathcal{Y}}f\,d\lambda=\mu(A)+E_\mu(g\,\mathbbm{1}_A)\,E_\nu(h)=\mu(A),
\\\int_{\mathcal{X}\times B}f\,d\lambda=\nu(B)+E_\mu(g)\,E_\nu(h\,\mathbbm{1}_B)=\nu(B).
\end{gather*}
Therefore, $f$ is a probability density with respect to $\lambda$ and $P_f\in\Gamma(\mu,\nu)$, that is, $f\in\mathcal{D}$. Similarly, one still obtains $f\in\mathcal{D}$ if $f$ is defined as
$$f(x,y)=1+\sum_{n=1}^\infty u_n\,g_n(x)\,h_n(y)$$
where each pair $(g_n,h_n)$ satisfies condition \eqref{bg67km} and the $u_n$ are constants such that $\sum_n\abs{u_n}\le 1$.

\medskip

\noindent Based on the previous remarks, to build a random density, it suffices to select $g_n$, $h_n$ and $u_n$ randomly. A straightforward way to do it is to let
\begin{gather}\label{r5b8j}
f(x,y)=1+\sum_{n=1}^\infty U_n\,g_n(x)\,h_n(y)
\end{gather}
where $g_n$ and $h_n$ are fixed deterministic functions satisfying condition \eqref{bg67km} while the $U_n$ are real random variables such that $\sum_n\abs{U_n}\le 1$. Then, for each $H\in\mathcal{H}$,
\begin{gather*}P_f(H)=\lambda(H)+\sum_{n=1}^\infty a_n(H)\,U_n\quad\quad\text{where }a_n(H)=\int_H g_n(x)\,h_n(y)\,\lambda(dx,dy).
\end{gather*}
Hence, the distribution of $P_f(H)$ is available at least when the $U_n$ are independent. In this case, in fact,
$$E\Bigl\{\exp\bigl(i\,t\,P_f(H)\bigr)\Bigr\}=\exp\bigl(i\,\lambda(H)\,t\bigr)\,\,\prod_{n=1}^\infty\,\phi_n\bigl(a_n(H)\,t\bigr)\quad\quad\text{for all }t\in\mathbb{R}$$
where $\phi_n$ is the characteristic function of $U_n$.

\medskip

\noindent In passing, we also note that, if $(\mathcal{X},\mathcal{F})=(\mathcal{Y},\mathcal{G})$ and $\mu=\nu$, it is possible to let $g_n=h_n$ for all $n$. In this case,
$$P_f(A\times B)=P_f(B\times A)\quad\quad\text{for all }A,\,B\in\mathcal{F}$$
so that the r.p.m. $P_f$ is exchangeable. Hence, if $\mu=\nu$ and $g_n=h_n$ for all $n$, one a.s. selects exchangeable elements of $\Gamma(\mu,\mu)$.

\medskip

\noindent The random density \eqref{r5b8j} is very simple but, just for this reason, potentially useful in real problems. We now turn to a more involved random density.

\medskip

\noindent Say that $F$ is a {\em random distribution function} on $\mathbb{R}^n$ if
$$F(x_1,\ldots,x_n)=P\Bigl((-\infty,x_1]\times\ldots\times (-\infty,x_n]\Bigr),\quad\quad (x_1,\ldots,x_n)\in\mathbb{R}^n,$$
for some r.p.m. $P$ on $(\mathbb{R}^n,\mathcal{B}(\mathbb{R}^n))$. Write $P_F$ to denote such a $P$ and $m$ to indicate the Lebesgue measure on $[0,1]$. A {\em random bivariate copula} is a random distribution function $C$ on $\mathbb{R}^2$ such that $P_C\in\Gamma(m,m)$ a.s.

\medskip

\noindent We now assume
$$\mathcal{X}=\mathcal{Y}=[0,1]\quad\text{and}\quad\mu=\nu=m.$$
However, the results obtained in this case can be easily extended to $\mathcal{X}=\mathcal{Y}=\mathbb{R}$. In fact, if $\mu$ and $\nu$ are Borel probabilities on $\mathbb{R}$ and $C$ is a random bivariate copula, then
$$F(x,y)=C\bigl[F_\mu(x),\,F_\nu(y)\bigr],\quad\quad (x,y)\in\mathbb{R}^2,$$
is a random distribution function on $\mathbb{R}^2$ with marginals $F_\mu$ and $F_\nu$, where
$$F_\mu(x)=\mu\bigl((-\infty,x]\bigr)\quad\text{and}\quad F_\nu(y)=\nu\bigl((-\infty,y]\bigr).$$
Hence, any random bivariate copula $C$ induces a random distribution function $F$ on $\mathbb{R}^2$ such that $P_F\in\Gamma(\mu,\nu)$. Equivalently, any prior on bivariate copulas induces a prior on $\Gamma(\mu,\nu)$.

\medskip

\noindent Let $\varphi:\mathbb{R}\rightarrow\mathbb{R}$ be a bounded Borel function and $W_i=\bigl\{W_i(t):t\in [0,1]\bigr\}$ a real stochastic process with Borel paths, where $i=1,2$. Suppose also that
\begin{gather}\label{b43z1m}
\sup_{t\in [0,1]}\,\,\Abs{\varphi\bigl(W_i(t)\bigr)-\int_0^1\varphi\bigl(W_i(s)\bigr)\,ds}\le 1\quad\quad\text{for }i=1,2.
\end{gather}
Under these conditions, define
$$g(t)=\varphi\bigl(W_1(t)\bigr)-\int_0^1\varphi\bigl(W_1(s)\bigr)\,ds\quad\text{and}\quad h(t)=\varphi\bigl(W_2(t)\bigr)-\int_0^1\varphi\bigl(W_2(s)\bigr)\,ds.$$
Our next result deals with the random density $f(x,y)=1+g(x)\,h(y)$ when $W_1$ and $W_2$ are standard Brownian motions. We denote by
$$L_i=\bigl\{L_i(t,x):t\in [0,1],\,\,x\in\mathbb{R}\bigr\}$$
the local time process corresponding to $W_i$. Here, $L_i$ is meant according to \cite[p. 201]{KARSH}. Among other things, $L_i$ has non-negative continuous paths and the map $x\mapsto 2L_i(t,x)$ is a density of the occupation measure of $W_i$ on $[0,t]$. The latter measure is defined as
$$\mathcal{O}_{i,t}(A)=\int_0^t \mathbbm{1}_A\bigl(W_i(s)\bigr)\,ds\quad\quad\text{for all }A\in\mathcal{B}(\mathbb{R}).$$

\medskip

\begin{theorem}
Let $W_1$ and $W_2$ be standard Brownian motions and $\varphi:\mathbb{R}\rightarrow\mathbb{R}$ a Borel function such that $-1/2\le\varphi\le 1/2$ or $0\le\varphi\le 1$. Define
\begin{gather*}
f(x,y)=1+\left\{\varphi\bigl(W_1(x)\bigr)-\int_0^1\varphi\bigl(W_1(s)\bigr)\,ds\right\}\,\left\{\varphi\bigl(W_2(y)\bigr)-\int_0^1\varphi\bigl(W_2(s)\bigr)\,ds\right\}
\end{gather*}
for all $(x,y)\in [0,1]^2$. Then, $f$ is a random density and
$$P_f\Bigl([0,a]\times [0,b]\Bigr)=ab\,+\,4\,U_1(a)\,U_2(b)$$
for all $0\le a,\,b\le 1$, where
\begin{gather*}
U_i(t)=\int_{-\infty}^\infty\bigl\{L_i(t,x)-t\,L_i(1,x)\bigr\}\,\varphi(x)\,dx\quad\quad\text{for }i=1,2\text{ and }t\in [0,1].
\end{gather*}
\end{theorem}

\begin{proof}
Condition \eqref{b43z1m} holds since $-1/2\le\varphi\le 1/2$ or $0\le\varphi\le 1$. Hence, $f$ is a random density. Let $\mathcal{O}_{i,t}$ be the occupation measure of $W_i$ on $[0,t]$. Since $2\,L_i(t,\cdot)$ is a density of $\mathcal{O}_{i,t}$, one obtains
$$\int_0^t\varphi\bigl(W_i(s)\bigr)\,ds=\int_{-\infty}^\infty\varphi(x)\,\mathcal{O}_{i,t}(dx)=2\,\int_{-\infty}^\infty L_i(t,x)\,\varphi(x)\,dx.$$
Therefore,
\begin{gather*}
P_f\bigl([0,a]\times [0,b]\bigr)=\int_0^a\int_0^b f(x,y)\,dxdy
\\=ab+\left\{\int_0^a\varphi\bigl(W_1(s)\bigr)\,ds-a\,\int_0^1\varphi\bigl(W_1(s)\bigr)\,ds\right\}\,\,\left\{\int_0^b\varphi\bigl(W_2(s)\bigr)\,ds-b\,\int_0^1\varphi\bigl(W_2(s)\bigr)\,ds\right\}
\\=ab+4\,\int_{-\infty}^\infty\bigl\{L_1(a,x)-a\,L_1(1,x)\bigr\}\,\varphi(x)\,dx\,\,\int_{-\infty}^\infty\bigl\{L_2(b,x)-b\,L_2(1,x)\bigr\}\,\varphi(x)\,dx.
\end{gather*}
\end{proof}

\medskip

\noindent The random variables $U_1(a)$ and $U_2(b)$ are independent whenever $W_1$ and $W_2$ are independent. In this case, to evaluate the distribution of $P_f\Bigl([0,a]\times [0,b]\Bigr)$, it suffices to know the distribution of $U_i(t)$. Possibly, the latter distribution is known, at least if $\varphi$ has a simple form, but we are not aware of this fact.

\medskip

\subsection{Second approach} We now develop a different idea to build a random density. We still assume $\mathcal{X}=\mathcal{Y}=[0,1]$ and $\mu=\nu=m$ where $m$ is the Lebesgue measure on $[0,1]$. As explained in Section \ref{bew5}, however, any result obtained in this case extends easily to $\mathcal{X}=\mathcal{Y}=\mathbb{R}$. Note also that $\lambda=\mu\times\nu=m\times m$ is Lebesgue measure on $[0,1]^2$.

\medskip

\noindent Fix an integer $k\ge 1$ and define the intervals
$$I_j=\left(\frac{j-1}{k},\,\frac{j}{k}\right]\quad\text{for }j=1,\ldots,k.$$
Call $\Sigma$ the set of all permutations $\sigma=\bigl(\sigma(1),\ldots,\sigma(k)\bigr)$ of $(1,\ldots,k)$. For each $\sigma\in\Sigma$, denote by $f_\sigma$ the uniform density (with respect to $\lambda$) on the set
$$S_\sigma=\bigcup_{j=1}^kI_j\times I_{\sigma(j)}.$$
Hence,
$$f_\sigma(x,y)=k\,\mathbbm{1}_{S_\sigma}(x,y)=k\,\sum_{j=1}^k\mathbbm{1}_{I_j}(x)\,\mathbbm{1}_{I_{\sigma(j)}}(y).$$
For any Borel set $A\subset [0,1]$,
\begin{gather*}
\int_{A\times [0,1]}f_\sigma\,d\lambda=k\,\lambda\Bigl(S_\sigma\cap\bigl(A\times [0,1]\bigr)\Bigr)=\sum_{j=1}^k m\bigl(A\cap I_j\bigr)=m(A),
\\\int_{[0,1]\times A}f_\sigma\,d\lambda=k\,\lambda\Bigl(S_\sigma\cap\bigl([0,1]\times A\bigr)\Bigr)=\sum_{j=1}^k m\bigl(A\cap I_{\sigma(j)}\bigr)=m(A).
\end{gather*}
Therefore, $P_{f_\sigma}\in\Gamma(m,m)$ or equivalently $f_\sigma\in\mathcal{D}$. Since $\mathcal{D}$ is a convex set, one also obtains
$$\sum_{\sigma\in\Sigma}u_\sigma\,f_\sigma\in\mathcal{D}$$
provided the $u_\sigma$ are non-negative constants such that $\sum_{\sigma\in\Sigma}u_\sigma=1$.

\medskip

\noindent Based on the previous remarks, to get a random density, it suffices to select the weights $u_\sigma$ in a random way. To do this, we write $\Sigma$ as
$$\Sigma=\bigl\{\sigma_1,\sigma_2,\ldots,\sigma_{k!}\bigr\}$$
and we fix a random vector $U=\bigl(U_1,\ldots,U_{k!}\bigl)$ satisfying $U_i\ge 0$ for all $i$ and $\sum_{i=1}^{k!}U_i=1$ a.s. In this notation, we focus on the random density
\begin{gather}\label{bj87f4w}
f=\sum_{i=1}^{k!}U_i\,f_{\sigma_i}.
\end{gather}

\medskip

\begin{theorem}\label{v5y7m}
If $f$ is defined by \eqref{bj87f4w}, then
$$P_f\Bigl([0,a]\times [0,b]\Bigr)=\sum_{i=1}^{k!}U_i\,\left\{\sum_{j=1}^{[ka]}m\Bigl(I_{\sigma_i(j)}\cap [0,b]\Bigr)+(ka-[ka])\,m\Bigl(I_{\sigma_i([ka]+1)}\cap [0,b]\Bigr)\right\}$$
for all $a,\,b\in [0,1]$. Moreover, $f$ can be written as
$$f(x,y)=k\,\sum_{j,h=1}^kD_{j,h}\,\mathbbm{1}_{I_j}(x)\,\mathbbm{1}_{I_h}(y)\quad\text{where}\quad D_{j,h}=\sum_{i:\sigma_i(j)=h}U_i.$$
\end{theorem}

\begin{proof}
Just note that
\begin{gather*}
P_f\Bigl([0,a]\times [0,b]\Bigr)=\sum_{i=1}^{k!}U_i\,P_{f_{\sigma_i}}\Bigl([0,a]\times [0,b]\Bigr)=k\,\sum_{i=1}^{k!}U_i\,\lambda\Bigl(S_{\sigma_i}\cap\bigl([0,a]\times [0,b]\bigr)\Bigr)
\\=k\,\sum_{i=1}^{k!}U_i\,\sum_{j=1}^k m\bigl(I_j\cap [0,a]\bigr)\,m\bigl(I_{\sigma_i(j)}\cap [0,b]\bigr)
\\=\sum_{i=1}^{k!}U_i\,\left\{\sum_{j=1}^{[ka]}m\Bigl(I_{\sigma_i(j)}\cap [0,b]\Bigr)+(ka-[ka])\,m\Bigl(I_{\sigma_i([ka]+1)}\cap [0,b]\Bigr)\right\}.
\end{gather*}
Similarly,
\begin{gather*}
\frac{1}{k}\,f(x,y)=\sum_{i=1}^{k!}U_i\,\sum_{j=1}^k\mathbbm{1}_{I_j}(x)\,\mathbbm{1}_{I_{\sigma_i(j)}}(y)=\sum_{j=1}^k\mathbbm{1}_{I_j}(x)\sum_{i=1}^{k!}U_i\,\mathbbm{1}_{I_{\sigma_i(j)}}(y)
\\=\sum_{j=1}^k\mathbbm{1}_{I_j}(x)\sum_{h=1}^k\mathbbm{1}_{I_h}(y)\,\sum_{i:\sigma_i(j)=h}U_i=\sum_{j,h=1}^kD_{j,h}\,\mathbbm{1}_{I_j}(x)\,\mathbbm{1}_{I_h}(y).
\end{gather*}
\end{proof}

\medskip

\noindent By Theorem \ref{v5y7m}, for simple choices of the vector $U$, the probability distribution of $P_f\Bigl([0,a]\times [0,b]\Bigr)$ can be written in closed form. For instance, one could select a proper subset of permutations, say $\{\sigma_1,\ldots,\sigma_m\}$, and define
$$U_{m+1}=\ldots=U_{k!}=0\quad\text{and}\quad (U_1,\ldots,U_m)\sim\text{Dir}(\alpha_1,\ldots,\alpha_m)$$
for some $\alpha_1,\ldots,\alpha_m>0$. Note also that, if $ka$ and $kb$ are both integers, Theorem \ref{v5y7m} yields
$$P_f\Bigl([0,a]\times [0,b]\Bigr)=\frac{1}{k}\,\sum_{i=1}^{k!}\text{card}\bigl\{j:1\le j\le ka\text{ and }\sigma_i(j)\le kb\bigr\}\,U_i.$$

\medskip

\noindent Finally, the random density $f$ could be generalized by replacing $k$ with an integer-valued random variable $K$ independent of $U$. We don't develop this hint but we note that, if $\mathbb{P}(K=k)>0$ for each $k\ge 1$, the probability distribution of $P_f$ has a very large support. To motivate the latter (vague) claim, we prove that any element of $\Gamma(m,m)$ can be weakly approximated by $P_g$ for some density $g\in\mathcal{D}$ of the form
\begin{gather}\label{v7u99x2}
g(x,y)=\sum_{j,h=1}^kd_{j,h}\,\mathbbm{1}_{I_j}(x)\,\mathbbm{1}_{I_h}(y)
\end{gather}
where the $d_{j,h}$ are suitable constants. In the next result, $d_{BL}$ denotes the {\em bounded Lipschitz metric} between Borel probabilities on $[0,1]$, that is
$$d_{BL}(p,q)=\sup_\phi\,\Abs{\int\phi\,dp-\int\phi\,dq}$$
where $p$ and $q$ are Borel probabilities on $[0,1]$ and $\sup$ is over the 1-Lipschitz functions $\phi:[0,1]\rightarrow [-1,1]$.

\medskip

\begin{theorem}
Given $k\ge 1$ and $p\in\Gamma(m,m)$, define $g$ by \eqref{v7u99x2} where
$$d_{j,h}=k^2\,p\bigl(I_j\times I_h\bigr).$$
Then,
$$g\in\mathcal{D}\quad\text{and}\quad d_{BL}(p,\,P_g)\le\frac{2\,\sqrt{2}}{k}.$$
\end{theorem}

\begin{proof}
Since $p\in\Gamma(m,m)$, it is straightforward to check that
$$P_g\bigl(A\times [0,1]\bigr)=P_g\bigl([0,1]\times A\bigr)=m(A)$$
for each Borel set $A\subset [0,1]$. Hence, $g\in\mathcal{D}$. Next, fix a 1-Lipschitz function $\phi:[0,1]\rightarrow [-1,1]$ and define
$$\alpha=\sum_{j,h=1}^k\phi\bigl(j/k,\,h/k\bigr)\,p\bigl(I_j\times I_h\bigr).$$
Since $\phi$ is 1-Lipschitz,
$$\abs{\phi(x,y)-\phi\bigl(j/k,\,h/k\bigr)}\le\norm{(x,y)-(j/k,\,h/k)}\le\frac{\sqrt{2}}{k}\quad\text{for all }(x,y)\in I_j\times I_h$$
where $\norm{\cdot}$ is the Euclidean nonrm on $\mathbb{R}^2$. Therefore,
\begin{gather*}
\Abs{\int\phi\,dp-\alpha}=\Abs{\sum_{j,h=1}^k\int_{I_j\times I_h}\bigl\{\phi(x,y)-\phi\bigl(j/k,\,h/k\bigr)\bigr\}\,p(dx,dy)}
\\\le\sum_{j,h=1}^k\,\sup_{(x,y)\in I_j\times I_h}\,\abs{\phi(x,y)-\phi\bigl(j/k,\,h/k\bigr)}\,p\bigl(I_j\times I_h\bigr)\le\frac{\sqrt{2}}{k}.
\end{gather*}
Since $P_g(I_j\times I_h)=p(I_j\times I_h)$ for all $j$ and $h$, one similarly obtains
$$\Abs{\int\phi\,dP_g-\alpha}\le\frac{\sqrt{2}}{k}.$$
Hence,
$$\Abs{\int\phi\,dp-\int\phi\,dP_g}\le\Abs{\int\phi\,dp-\alpha}+\Abs{\alpha-\int\phi\,dP_g}\le\frac{2\,\sqrt{2}}{k}.$$
\end{proof}

\medskip

\section{Dirichlet type priors on $\Gamma(\mu)$}

\noindent To build a prior on $\Gamma(\mu)$, a naive idea is to fix a r.p.m. $Q$ on $(\mathcal{Y},\mathcal{G})$ and to let
$$P=\mu\times Q.$$
Then, $P$ is a r.p.m. on $(\mathcal{X}\times\mathcal{Y},\,\mathcal{F}\otimes\mathcal{G})$ and $P(A\times\mathcal{Y})=\mu(A)$ for all $A\in\mathcal{F}$, so that the probability distribution $\Pi$ of $P$ satisfies $\Pi\bigl(\Gamma(\mu)\bigr)=1$. It is tempting to take $Q$ a Dirichlet r.p.m. In this case, $P$ is basically a mixture of Dirichlet r.p.m.'s as defined in \cite{ANTONIAK}. The main features of $P$ are collected in the next result. To state it, for all $x\in\mathcal{X}$ and $H\subset\mathcal{X}\times\mathcal{Y}$, we denote by
$$H^x=\bigl\{y\in\mathcal{Y}:(x,y)\in H\bigr\}$$
the section of $H$ with respect to $x$.

\begin{theorem}\label{4xg7y}
Let $P=\mu\times Q$, where $Q$ is a Dirichlet r.p.m. on $(\mathcal{Y},\mathcal{G})$ with parameters $c$ and $\nu$ (as defined in Section \ref{b78m2}). Moreover, let $Z_n=(X_n,Y_n)$, $n\ge 1$, be an exchangeable sequence such that
$$(Z_n)\mid P\,\text{ is i.i.d. with }\,Z_1\sim P.$$
Then:

\medskip

\begin{itemize}

\item[(a)] For each finite measurable partition $H_1,\ldots,H_m$  of $\mathcal{X}\times\mathcal{Y}$,
$$\bigl(P(H_1),\ldots,P(H_m)\bigr)\,\sim\,\int_\mathcal{X}\text{Dir}\bigl[c\,\nu(H_1^x),\ldots,c\,\nu(H_m^x)\bigr]\,\mu(dx);$$

\medskip

\item[(b)] For each $n\ge 1$, conditionally on $(Z_1,\ldots,Z_n)$, one obtains
\begin{gather*}
\bigl(P(H_1),\ldots,P(H_m)\bigr)\,\sim\,\int_\mathcal{X}\text{Dir}\bigl[\nu_n(H_1^x),\ldots,\nu_n(H_m^x)\bigr]\,\mu(dx)
\end{gather*}
where $H_1,\ldots,H_m$ is a finite measurable partition of $\mathcal{X}\times\mathcal{Y}$ and $\nu_n$ is the random measure
$$\nu_n=c\,\nu+\sum_{i=1}^n\delta_{Y_i};$$

\medskip

\item[(c)] For all $A\in\mathcal{F}$ and $B\in\mathcal{G}$,
$$\mathbb{P}(X_{n+1}\in A,\,Y_{n+1}\in B\mid Z_1,\ldots,Z_n)=\mu(A)\,\,\frac{\nu_n(B)}{c+n}\quad\quad\text{a.s.}$$

\end{itemize}

\end{theorem}

\begin{proof}
First note that $H_1^x,\ldots,H_m^x$ is a finite measurable partition of $\mathcal{Y}$ for each $x\in\mathcal{X}$. Since $Q$ is Dirichlet with parameters $c$ and $\nu$, it follows that
$$
\bigl(Q(H_1^x),\ldots,Q(H_m^x)\bigr)\,\sim\,\text{Dir}\bigl[c\,\nu(H_1^x),\ldots,c\,\nu(H_m^x)\bigr].
$$
Moreover, $P(H)=\int_\mathcal{X}Q(H^x)\mu(dx)$ for each $H\in\mathcal{F}\otimes\mathcal{G}$. Hence,
\begin{gather*}
\bigl(P(H_1),\ldots,P(H_m)\bigr)=\left(\int_\mathcal{X}Q(H_1^x)\,\mu(dx),\ldots,\int_\mathcal{X}Q(H_m^x)\,\mu(dx)\right)
\\=\int_\mathcal{X}\bigl(Q(H_1^x),\ldots,Q(H_m^x)\bigr)\,\mu(dx)\,\sim\,\int_\mathcal{X}\text{Dir}\bigl[c\,\nu(H_1^x),\ldots,c\,\nu(H_m^x)\bigr]\,\mu(dx).
\end{gather*}
This proves point (a). As to (b), it suffices noting that:

\medskip

$-$ Conditionally on $(Y_1,\ldots,Y_n)$, the r.p.m. $Q$ is Dirichlet with parameters $(c+n)$ and $\frac{\nu_n}{c+n}$; Hence, conditionally on $(Y_1,\ldots,Y_n)$, one obtains

$$\bigl(Q(H_1^x),\ldots,Q(H_m^x)\bigr)\,\sim\,\text{Dir}\bigl[\nu_n(H_1^x),\ldots,\nu_n(H_m^x)\bigr];$$

\medskip

$-$ $P$ is conditionally independent of $(X_1,\ldots,X_n)$ given $(Y_1,\ldots,Y_n)$.

\medskip

\noindent Taking these two facts into account, point (b) can be proved exactly as (a). Finally, to get (c), consider a sequence
$$Z_n^*=(X_n^*,Y_n^*)$$
of random variables with values in $\mathcal{X}\times\mathcal{Y}$ and predictive distributions
$$\mathbb{P}(Z_{n+1}^*\in\cdot\mid Z_1^*,\ldots,Z_n^*)=\frac{c\,(\mu\times\nu)+\sum_{i=1}^n(\mu\times\delta_{Y_i^*})}{c+n}.$$
By the results of \cite{BDLPR23} (see Theorems 7 and 9), $(Z_n^*)$ is exchangeable and has the same prior as $(Z_n)$. (That is, the prior of both $(Z_n^*)$ and $(Z_n)$ is the probability distribution of $P$). Therefore, $(Z_n)\sim (Z_n^*)$, which in turn implies
$$\mathbb{P}(X_{n+1}\in A,\,Y_{n+1}\in B\mid Z_1,\ldots,Z_n)=\frac{c\,\mu(A)\,\nu(B)+\mu(A)\,\sum_{i=1}^n\delta_{Y_i}(B)}{c+n}=\mu(A)\,\,\frac{\nu_n(B)}{c+n}.$$
\end{proof}

\medskip

\noindent The idea underlying Theorem \ref{4xg7y} can be developed further. Suppose
$$\mathcal{X}=\mathcal{Y}=\mathbb{R}$$
and denote again by $F_\mu(t)=\mu\bigl((-\infty,t]\bigr)$ and $F_\nu(t)=\nu\bigl((-\infty,t]\bigr)$ the distribution functions corresponding to $\mu$ and $\nu$. Moreover, fix a bivariate copula $C$, a random distribution function $G$ on $\mathbb{R}$, and define
\begin{gather}\label{v67u8x2}
F(x,y)=C\bigl[F_\mu(x),\,G(y)\bigr]\quad\quad\text{for all }(x,y)\in\mathbb{R}^2.
\end{gather}
Then, $F$ is a random distribution function on $\mathbb{R}^2$ and $P_F\in\Gamma(\mu)$ by construction. (Recall that $P_F$ denotes the r.p.m. induced by $F$). If $C$ is the product copula, equation \eqref{v67u8x2} yields $F(x,y)=F_\mu(x)\,G(y)$ so that $P_F=\mu\times P_G$. Hence, in the special case where $C$ is the product copula and $P_G$ is Dirichlet, $P_F$ reduces to the r.p.m. involved in Theorem \ref{4xg7y}.

\medskip

\begin{theorem}\label{v56j9}
Define $F$ by \eqref{v67u8x2} where $P_G$ is Dirichlet with parameters $c$ and $\nu$. Fix $(x,y)\in\mathbb{R}^2$ such that $0<F_\mu(x),\,F_\nu(y)<1$ and denote by $\mathbb{B}_y$ the beta distribution function with parameters $c\,F_\nu(y)$ and $c\,(1-F_\nu(y))$. Then,
\begin{gather*}
\mathbb{P}\Bigl(F(x,y)\le a\Bigr)=\mathbb{B}_y\bigl[r(x,a)\bigr]\quad\quad\text{for all }a\in [0,\,F_\mu(x)]
\end{gather*}
where
$$r(x,a)=\sup\bigl\{v\in [0,1]:\,C\bigl[F_\mu(x),\,v\bigr]=a\bigr\}.$$
Moreover, if $Z_n=(X_n,Y_n)$, $n\ge 1$, is an exchangeable sequence such that
$$(Z_n)\mid F\,\text{ is i.i.d. with }\,Z_1\sim F,$$
then
\begin{gather*}
\mathbb{P}\Bigl(F(x,y)\le a\mid Y_1,\ldots,Y_n\Bigr)=\mathbb{B}_{n,y}\bigl[r(x,a)\bigr]\quad\quad\text{a.s. for all }n\ge 1,
\end{gather*}
where $\mathbb{B}_{n,y}$ is a beta random distribution function with parameters
$$c\,F_\nu(y)+\sum_{i=1}^n\mathbbm{1}_{\{Y_i\le y\}}\quad\text{and}\quad c\,(1-F_\nu(y))+\sum_{i=1}^n\mathbbm{1}_{\{Y_i> y\}}.$$
\end{theorem}

\begin{proof}
For each $v\in [0,1]$, define
$$\phi_x(v)=C\bigl[F_\mu(x),\,v\bigr].$$
Then, $\phi_x$ is a non-decreasing continuous map from $[0,1]$ onto $[0,\,F_\mu(x)]$. Moreover,
$$\bigl\{F(x,y)\le a\bigr\}=\bigl\{\phi_x(G(y))\le a\bigr\}=\bigl\{G(y)\le r(x,a)\bigr\}\quad\quad\text{for each }a\in [0,\,F_\mu(x)].$$
Since $P_G$ is Dirichlet with parameters $c$ and $\nu$, then $G(y)$ has a beta distribution with parameters $c\,F_\nu(y)$ and $c\,(1-F_\nu(y))$. Therefore,
$$\mathbb{P}\Bigl(F(x,y)\le a\Bigr)=\mathbb{P}\Bigl(G(y)\le r(x,a)\Bigr)=\mathbb{B}_y\bigl[r(x,a)\bigr]\quad\quad\text{for each }a\in [0,\,F_\mu(x)].$$
Next, conditionally on $(Y_1,\ldots,Y_n)$, $P_G$ is Dirichlet with parameters $(n+c)$ and $\frac{\nu_n}{n+c}$ where $\nu_n=c\,\nu+\sum_{i=1}^n\delta_{Y_i}$. Hence, $G(y)$ has a beta distribution with parameters
$$\nu_n\bigl((-\infty,y]\bigr)=c\,F_\nu(y)+\sum_{i=1}^n\mathbbm{1}_{\{Y_i\le y\}}\quad\text{and}\quad\nu_n\bigl((y,\infty)\bigr)=c\,(1-F_\nu(y))+\sum_{i=1}^n\mathbbm{1}_{\{Y_i> y\}}.$$
Therefore,
$$\mathbb{P}\Bigl(F(x,y)\le a\mid Y_1,\ldots,Y_n\Bigr)=\mathbb{P}\Bigl(G(y)\le r(x,a)\mid Y_1,\ldots,Y_n\Bigr)=\mathbb{B}_{n,y}\bigl[r(x,a)\bigr]\quad\text{a.s.}$$
\end{proof}

\medskip

\noindent It is worth noting that Theorem \ref{v56j9} provides the conditional distribution of $F$ given $(Y_1,\ldots,Y_n)$ but not the conditional distribution of $F$ given $(Z_1,\ldots,Z_n)$. In fact, we did not find any (useful) general formula for the latter.

\medskip

\noindent In Theorem \ref{v56j9}, $C$ is a {\em fixed} bivariate copula. An obvious development is to allow $C$ to be random. For instance, $C$ could be distributed according to Section \ref{ga332c}. Anyway, if $C$ is random, $r(x,a)$ is random as well (for it depends on $C$). To make this fact explicit, in our last result, we write $r_C(x,a)$ instead of $r(x,a)$.

\medskip

\begin{corollary}
Define $F$ by \eqref{v67u8x2} where $P_G$ is Dirichlet with parameters $c$ and $\nu$ and $C$ is a random bivariate copula. Suppose $C$ independent of $G$ and fix $(x,y)\in\mathbb{R}^2$ such that $0<F_\mu(x),\,F_\nu(y)<1$. Then,
\begin{gather*}
\mathbb{P}\Bigl(F(x,y)\le a\Bigr)=\int\mathbb{B}_y\bigl[r_C(x,a)\bigr]\,\Pi(dC)\quad\quad\text{for all }a\in [0,\,F_\mu(x)]
\end{gather*}
where $\Pi$ denotes the probability distribution of $C$.
\end{corollary}

\begin{proof}
Since $C$ and $G$ are independent, it suffices to apply Theorem \ref{v56j9}.
\end{proof}

\medskip

\noindent A last remark is that the results of this section (suitably adapted) are still true if the Dirichlet is replaced by any r.p.m. with a known probability distribution.

\end{document}